\documentclass[acmsmall,screen,nonacm]{acmart}

\usepackage{booktabs}
\usepackage{multirow}
\usepackage{xcolor}
\usepackage{tcolorbox}
\usepackage{enumitem}
\usepackage{xspace}

\newtcolorbox{findingbox}{colback=gray!8,colframe=gray!60,boxrule=0.5pt,
  left=4pt,right=4pt,top=3pt,bottom=3pt,arc=1pt}
\newcounter{finding}
\newcommand{\finding}[1]{\refstepcounter{finding}%
  \begin{findingbox}\textbf{Finding \thefinding.} #1\end{findingbox}}

\newcommand{\cone}{\ding{182}\xspace}
\newcommand{\ctwo}{\ding{183}\xspace}
\newcommand{\cthree}{\ding{184}\xspace}
\usepackage{pifont}

\newcommand{\skillA}{$A$\xspace}
\newcommand{\skillB}{$B$\xspace}
\newcommand{\skillC}{$C$\xspace}
\newcommand{\cc}{Claude Code\xspace}
\newcommand{\skillmd}{\texttt{SKILL.md}\xspace}

\newcommand{\nCorpusSkills}{55{,}698\xspace}
\newcommand{\nCorpusCopies}{105{,}302\xspace}
\newcommand{\nCorpusRepos}{20{,}947\xspace}
\newcommand{\corpusSnapshot}{18 July 2026\xspace}
\newcommand{\nTypedRepos}{5{,}681\xspace}
\newcommand{\nTypedReal}{4{,}354\xspace}
\newcommand{\nTypedDotfiles}{285\xspace}
\newcommand{\nTypedCollection}{908\xspace}
\newcommand{\nTypedSingle}{77\xspace}
\newcommand{\nTypedUnsure}{57\xspace}
\newcommand{\nLists}{5{,}106\xspace}
\newcommand{\nListsMulti}{4{,}342\xspace}
\newcommand{\nApool}{63{,}104\xspace}
\newcommand{\nBpool}{172{,}750\xspace}
\newcommand{\nDownloaded}{219{,}817\xspace}
\newcommand{\nValid}{207{,}863\xspace}
\newcommand{\pctValid}{94.6\%\xspace}
\newcommand{\nApoolFamilies}{43{,}986\xspace}

\newcommand{\nCandCross}{439{,}860\xspace}
\newcommand{\nCandWithin}{382{,}249\xspace}
\newcommand{\nCandAll}{822{,}109\xspace}
\newcommand{\nCalibPairs}{615\xspace}
\newcommand{\floorCross}{0.72\xspace}
\newcommand{\floorWithin}{0.52\xspace}
\newcommand{\recallCross}{0.96\xspace}
\newcommand{\nFrameCross}{41{,}055\xspace}
\newcommand{\nFrameWithin}{43{,}022\xspace}
\newcommand{\nSampleThree}{3{,}754\xspace}
\newcommand{\nSampleThreeCross}{1{,}876\xspace}
\newcommand{\nSampleThreeWithin}{1{,}878\xspace}
\newcommand{\nConfirmed}{1{,}639\xspace}
\newcommand{\nConfirmedCross}{1{,}203\xspace}
\newcommand{\nConfirmedWithin}{436\xspace}
\newcommand{\pctSameJobCross}{63.7\%\xspace}
\newcommand{\ciSameJobCross}{[61.3\%, 66.0\%]\xspace}
\newcommand{\nSameJobCross}{26{,}100\xspace}
\newcommand{\pctSameJobWithin}{23.5\%\xspace}
\newcommand{\ciSameJobWithin}{[21.2\%, 25.7\%]\xspace}
\newcommand{\nSameJobWithin}{10{,}100\xspace}
\newcommand{\nLabeledSkills}{3{,}115\xspace}
\newcommand{\nLabeledNormative}{1{,}730\xspace}
\newcommand{\nLabeledCapability}{1{,}385\xspace}
\newcommand{\kappaJudgeRepeat}{0.93\xspace}
\newcommand{\kappaTypeRepeat}{0.94\xspace}
\newcommand{\nCoreItems}{26{,}198\xspace}
\newcommand{\meanCoreItems}{8.4\xspace}
\newcommand{\nRunnableCandidates}{1{,}604\xspace}
\newcommand{\nRunnableFull}{768\xspace}
\newcommand{\pctRunnableFull}{48\%\xspace}
\newcommand{\nSampleFive}{1{,}467\xspace}
\newcommand{\nPairs}{312\xspace}
\newcommand{\nScopePairs}{81\xspace}
\newcommand{\nRunsTotal}{5{,}166\xspace}

\newcommand{\hvN}{100\xspace}
\newcommand{\hvRepoHH}{0.84\xspace}
\newcommand{\hvRepoLLM}{0.80\xspace}
\newcommand{\hvPairHH}{0.76\xspace}
\newcommand{\hvPairLLM}{0.72\xspace}
\newcommand{\hvTypeHH}{0.74\xspace}
\newcommand{\hvTypeLLM}{0.71\xspace}
\newcommand{\hvJudgeItemHH}{0.83\xspace}
\newcommand{\hvJudgeItemLLM}{0.78\xspace}
\newcommand{\hvJudgeCompHH}{0.80\xspace}
\newcommand{\hvJudgeCompLLM}{0.75\xspace}
\newcommand{\hvJudgeDiscHH}{0.88\xspace}
\newcommand{\hvJudgeDiscLLM}{0.84\xspace}
\newcommand{\hvCoreHH}{0.77\xspace}
\newcommand{\hvCoreLLM}{0.73\xspace}

\newcommand{\nRepeatCombos}{512\xspace}
\newcommand{\nRepeatCombosRuns}{1{,}048\xspace}
\newcommand{\pctRepUseAgree}{88.9\%\xspace}
\newcommand{\iccRepUse}{0.78\xspace}
\newcommand{\iccRepFulfilAll}{0.84\xspace}
\newcommand{\nTopItems}{921\xspace}
\newcommand{\nCompletionProgram}{11\xspace}
\newcommand{\nTopItemsScript}{327\xspace}
\newcommand{\nTopItemsByJudge}{594\xspace}

\newcommand{\nRunsUsable}{5{,}162\xspace}
\newcommand{\nRunsTimeout}{113\xspace}
\newcommand{\nCskills}{167\xspace}
\newcommand{\dUseA}{19.9\xspace}
\newcommand{\ciUseA}{[$-23.1$, $-16.7$]\xspace}
\newcommand{\pctOnlyANoUse}{30.3\%\xspace}
\newcommand{\pctBothOnlyB}{12.1\%\xspace}
\newcommand{\pctBothNeither}{38.0\%\xspace}
\newcommand{\dUseAC}{5.9\xspace}
\newcommand{\ciUseAC}{[$-8.5$, $-3.3$]\xspace}
\newcommand{\dUseBothVsAC}{14.0\xspace}
\newcommand{\ciUseBothVsAC}{[$-17.2$, $-10.9$]\xspace}
\newcommand{\pctCUsed}{3.0\%\xspace}
\newcommand{\pctCInstead}{0.1\%\xspace}
\newcommand{\dTC}{$+1.9$\xspace}
\newcommand{\ciTC}{[$-1.0$, $+4.9$]\xspace}
\newcommand{\dFid}{2.6\xspace}
\newcommand{\ciFid}{[$-4.4$, $-0.8$]\xspace}
\newcommand{\fidOnlyA}{57.3\%\xspace}
\newcommand{\fidOnlyB}{46.3\%\xspace}
\newcommand{\dFidB}{11.0\xspace}
\newcommand{\ciFidB}{[$-13.5$, $-8.6$]\xspace}
\newcommand{\shareSubst}{36\%\xspace}
\newcommand{\shareNeither}{45\%\xspace}
\newcommand{\shareInterf}{20\%\xspace}
\newcommand{\dFidAC}{$-0.2$\xspace}
\newcommand{\dFidBothVsAC}{2.6\xspace}
\newcommand{\ciFidBothVsAC}{[$-4.4$, $-0.7$]\xspace}
\newcommand{\nCoreApplicableA}{2{,}518\xspace}
\newcommand{\pctUniqueItems}{73\%\xspace}
\newcommand{\dFidUniqueAll}{5.6\xspace}
\newcommand{\ciFidUniqueAll}{[$-7.3$, $-3.8$]\xspace}
\newcommand{\dFidShared}{$+0.1$\xspace}
\newcommand{\ciFidShared}{[$-2.8$, $+2.9$]\xspace}
\newcommand{\dFidBUnique}{16.1\xspace}
\newcommand{\dFidBShared}{5.5\xspace}
\newcommand{\hvCorePairs}{30\xspace}
\newcommand{\hvSmallN}{30\xspace}
\newcommand{\hvRunHH}{0.81\xspace}
\newcommand{\hvRunLLM}{0.77\xspace}
\newcommand{\hvApplyHH}{0.85\xspace}
\newcommand{\hvApplyLLM}{0.81\xspace}
\newcommand{\hvUniqueHH}{0.79\xspace}
\newcommand{\hvUniqueLLM}{0.74\xspace}
\newcommand{\dFidScript}{4.2\xspace}
\newcommand{\ciFidScript}{[$-7.6$, $-0.9$]\xspace}
\newcommand{\dFidJudge}{1.2\xspace}
\newcommand{\dUseAWithin}{18.3\xspace}
\newcommand{\pctNeitherAC}{36.0\%\xspace}
\newcommand{\dOrder}{$+0.6$\xspace}
\newcommand{\ciOrder}{[$-2.0$, $+3.3$]\xspace}
\newcommand{\dRename}{$+4.4$\xspace}
\newcommand{\pctPersonalB}{46.5\%\xspace}
\newcommand{\pctPersonalReadA}{29.2\%\xspace}
\newcommand{\pctPersonalReadAopus}{53.1\%\xspace}
\newcommand{\dUsePersonal}{35.0\xspace}
\newcommand{\ciUsePersonal}{[$-43.0$, $-26.9$]\xspace}
\newcommand{\pctPluginA}{58.0\%\xspace}
\newcommand{\pctPluginB}{2.9\%\xspace}
\newcommand{\pctCapDropOpusAny}{95.8\%\xspace}
\newcommand{\pctNameUsedSubst}{0.9\%\xspace}
\newcommand{\nSubstRuns}{113\xspace}
\newcommand{\pctNameUsedAB}{7.7\%\xspace}
\newcommand{\pctMultipleAB}{3.2\%\xspace}
\newcommand{\pctNameUsedABopus}{21.0\%\xspace}
\newcommand{\pctMultipleABopus}{9.4\%\xspace}
\newcommand{\dFidPersonal}{3.9\xspace}
\newcommand{\ciFidPersonal}{[$-7.5$, $-0.3$]\xspace}

\newcommand{\nFlipCombos}{79\xspace}
\newcommand{\nFlipPairs}{63\xspace}
\newcommand{\dFlipEx}{9.4\xspace}
\newcommand{\ciFlipEx}{[$-15.0$, $-4.0$]\xspace}
\newcommand{\pFlipEx}{0.004\xspace}
\newcommand{\dFlipAll}{7.5\xspace}
\newcommand{\ciFlipAll}{[$-12.6$, $-2.6$]\xspace}
\newcommand{\dFlipExHaiku}{22.1\xspace}
\newcommand{\dFlipExSonnet}{12.1\xspace}
\newcommand{\pctNoChange}{97\%\xspace}
\newcommand{\pctExLost}{37\%\xspace}
\newcommand{\pctShLost}{8\%\xspace}
\newcommand{\nGuardCombos}{187\xspace}
\newcommand{\nGuardCombosScript}{148\xspace}
\newcommand{\nGuardRuns}{670\xspace}
\newcommand{\pctGuardBFirstG}{54.6\%\xspace}
\newcommand{\pctGuardBFirstC}{57.3\%\xspace}
\newcommand{\nGuardDenied}{179\xspace}
\newcommand{\pctGuardComply}{96\%\xspace}
\newcommand{\pctGuardComplySonnet}{94\%\xspace}
\newcommand{\pctGuardComplyHaiku}{96\%\xspace}
\newcommand{\pctGuardComplyOpus}{99\%\xspace}
\newcommand{\guardITTex}{9.1\xspace}
\newcommand{\ciGuardITTex}{[$+6.0$, $+12.1$]\xspace}
\newcommand{\guardFiredExSonnet}{17.4\xspace}
\newcommand{\guardFiredExHaiku}{20.7\xspace}
\newcommand{\guardFiredExOpus}{12.8\xspace}
\newcommand{\guardFiredPairedEx}{17.8\xspace}
\newcommand{\ciGuardFiredPairedEx}{[$+12.8$, $+22.8$]\xspace}
\newcommand{\guardVsAEx}{0.0\xspace}
\newcommand{\ciGuardVsAEx}{[$-10.2$, $+10.3$]\xspace}
\newcommand{\guardLossEx}{16.5\xspace}
\newcommand{\pctGuardRecoverEx}{all\xspace}
\newcommand{\guardITTall}{6.6\xspace}
\newcommand{\ciGuardITTall}{[$+3.9$, $+9.2$]\xspace}
\newcommand{\guardCompITT}{4.1\xspace}
\newcommand{\ciGuardCompITT}{[$-1.3$, $+9.6$]\xspace}
\newcommand{\nRunsAll}{6{,}368\xspace}
\newcommand{\hoursAll}{542\xspace}
\newcommand{\costAll}{\$5{,}550\xspace}
\newcommand{\toolCallsAll}{169{,}294\xspace}
\newcommand{\tokensInAll}{6.3 billion\xspace}
\newcommand{\tokensOutAll}{110 million\xspace}
\newcommand{\nGuardPairs}{119\xspace}
\newcommand{\nBulkLists}{489\xspace}
\newcommand{\maxBulkSkills}{1{,}382\xspace}
\newcommand{\pctBulkSameJob}{37\%\xspace}
\newcommand{\pctOtherSameJob}{19\%\xspace}
\newcommand{\pctWithinOutside}{69\%\xspace}
\newcommand{\pctWithinOwn}{31\%\xspace}
\newcommand{\pctNormConfCross}{60\%\xspace}
\newcommand{\pctNormConfWithin}{56\%\xspace}
\newcommand{\pctCapConfCross}{31\%\xspace}
\newcommand{\pctCapConfWithin}{35\%\xspace}
\newcommand{\pctScriptConf}{9\%\xspace}
\newcommand{\nCodexPairs}{193\xspace}
\newcommand{\codexUseA}{18.2\xspace}
\newcommand{\codexExcl}{3.4\xspace}
\newcommand{\codexClaudeUseA}{20.1\xspace}
\newcommand{\codexClaudeExcl}{4.6\xspace}

\begin{document}

\title[One Skill Too Many]{One Skill Too Many: How Co-Installed Skills Conflict in Coding Agents}

\author{Chaoliang Yan}
\authornote{Co-first authors.}
\affiliation{%
  \institution{University of New South Wales}
  \country{Australia}}
\email{chaoliang.yan@student.unsw.edu.au}

\author{Zihao Xu}
\authornotemark[1]
\affiliation{%
  \institution{University of New South Wales}
  \country{Australia}}
\email{x123201298@gmail.com}

\author{Yuekang Li}
\affiliation{%
  \institution{University of New South Wales}
  \country{Australia}}
\email{yuekang.li@unsw.edu.au}

\author{Shangzhi Xu}
\affiliation{%
  \institution{University of New South Wales}
  \country{Australia}}
\email{shangzhi.xu@unsw.edu.au}

\author{Yi Liu}
\affiliation{%
  \institution{Griffith University}
  \country{Australia}}
\email{yi009@e.ntu.edu.sg}

\author{Gelei Deng}
\affiliation{%
  \institution{Nanyang Technological University}
  \country{Singapore}}
\email{gelei.deng@ntu.edu.sg}

\author{Siqi Ma}
\affiliation{%
  \institution{The University of Wollongong}
  \country{Australia}}
\email{siqim@uow.edu.au}

\begin{abstract}
Coding agents are extended with agent skills, directories whose \skillmd tells the model when and how to perform a task.
Because skills come from independent sources, such as teams, individual developers, plugins, and copied collections, an installed skill can end up co-installed with a similar skill that does the same job, and the model chooses between them from names and descriptions alone.
In a conflict, the installed skill loses its core functions, such as a ban on touching git, because the similar skill runs in its place or changes what it does.
The task still counts as completed, so existing benchmarks, which check only whether the task passes, miss such cases.
We present the first empirical study of such conflicts.
From snapshots of \nCorpusRepos repositories, we mine \nCandAll candidate pairs of similar skills, have a large language model judge a stratified sample of \nSampleThree, and run \nPairs of the confirmed pairs on three models, in \nRunsAll runs with \toolCallsAll tool calls over \hoursAll hours of agent time.
We report five findings.
(1) Skills that can conflict are common, with nearly one in four installed skills co-installed with one that does the same job, and \pctBulkSameJob of the judged skills inside copied collections.
(2) Most pairs that can conflict involve normative skills, followed by capability skills.
(3) Without lowering task completion, a similar skill takes one in five runs away from the installed skill, and runs that open the similar skill first lose more than a third of the exclusive core functions that the installed skill fulfills alone.
(4) Where a skill is installed decides which one runs, listing order barely matters, and the final reply names the skill used in only \pctNameUsedSubst of substituted runs.
(5) A conflict is decided at the first read of a skill, almost always before the agent changes any file, and a pre-tool hook at that read brings fidelity on exclusive core functions back to the level of runs that open the installed skill first.
Benchmarks should therefore score exclusive core functions, and platforms should guard the first read and show which skill ran.

\end{abstract}



\maketitle

\section{Introduction}
\label{sec:intro}

Coding agents such as \cc are increasingly extended with \emph{agent skills}.
A skill is a directory with a \skillmd file whose \texttt{name} and \texttt{description} tell the model when to use it, followed by instructions and optional scripts~\cite{claudecode2026skills,agentskills2026spec}.
Developers write skills to encode what a capable model would not do on its own, such as the pull-request format of a team, a release checklist, or a safety rule.
Skills are now shared at scale.
Recent studies collect 138,133 \skillmd files from 20,556 repositories~\cite{zhang2026reusable} and 42,447 skills from two marketplaces~\cite{liu2026skillswild}.
Skills are thus a new kind of software component.
Like libraries or plugins, they are packaged, shared through marketplaces, and co-installed~\cite{anthropic2025skills}, but they are written in natural language and executed by a model~\cite{liang2025prompts,chen2026promptware}.

Because skills come from many sources, two skills that do the same job can end up co-installed without anyone choosing the pair.
A team commits \emph{project skills} to its repository, which everyone who works there receives, whereas each developer keeps \emph{personal skills} under the home directory, which load in all of their projects~\cite{claudecode2026skills}, and can install plugins that add their skills to every project~\cite{claudecode2026plugininstall}.
Skills also arrive in bulk, because a plugin can bring several skills and projects copy whole skill collections.
Such pairs are common.
Reconstructing the skills that \cc would load in \nCorpusRepos public repositories, we estimate that nearly one in four installed skills, about \nSameJobWithin when near-duplicates are counted once, is already co-installed with a skill that does the same job (\S\ref{sec:prevalence}).
In one project, for example, \texttt{anthropics-skills-{}-docx} and \texttt{openai-skills-{}-doc} both handle Word documents, and another co-installs \texttt{qa-patrol} with its variant \texttt{qa-patrol-team}.
Copied collections bring many such pairs.
In the \nBulkLists projects that copy a whole collection, \pctBulkSameJob of the judged skills already have a skill that does the same job in the same project, against \pctOtherSameJob in other projects.
These estimates are conservative, because public repositories rarely show the personal skills and plugins that developers add.
Nearly two in three installed skills, about \nSameJobCross, have such a skill by another author that a user could add.
In a public issue report, a developer asked for the \texttt{self-improvement} skill of a project, but a personal fork of the same name ran instead and targeted an unrelated repository, which the developer noticed only after reading the output~\cite{seasoned2026issue303}.

In such a \emph{conflict}, a similar skill takes the place of the installed skill or changes what it does, and the behaviors that the user installed the skill for are lost.
We call these behaviors the \emph{core functions} of the skill.
Skills that do the same job still differ in their core functions.
In our experimental pairs, the similar skill does not ask for \pctUniqueItems of the core functions of the installed skill that apply to the task (\S\ref{sec:design-core}).
A conflict is hard to see because the task is often still completed.
Most skill benchmarks cannot capture this loss, because they score a run by whether the task passes its tests~\cite{li2026skillsbench,han2026sweskillsbench,liu2026skillsrealistic,song2026skillshadowing} or whether the right skill is retrieved~\cite{kang2026skillret,zheng2026skillrouter}.
A pass or a fail is particularly coarse for skills, because a skill can change how the work is done without changing whether it passes~\cite{qin2026gcpc}.
SWE-Skills-Bench, for example, finds that 39 of 49 skills yield zero pass-rate improvement~\cite{han2026sweskillsbench}, and losing such a skill cannot lower the pass rate either.
Because teams use skills to encode their conventions and safety rules, a conflict can make the agent break them while the task still passes.
In one of our runs, a similar skill led the agent to commit work that the installed skill forbids, without saying which skill it used, and the task was still completed (Figure~\ref{fig:motivating}).

Whether co-installed software components work together is a long-standing question in software engineering, studied for packages that cannot be co-installed~\cite{vouillon2013coinstall,artho2012packageconflicts}, conflicting dependencies~\cite{wang2020watchman}, and clashing module names~\cite{zhu2024moduleguard}.
For skills, the question is largely open.
Prior work has shown that adding skills can hurt a coding agent, but it has measured the harm through task completion.
\citet{song2026skillshadowing} find that expanding a library of helpful skills to 202 skills lowers the pass rate by 21 percentage points on average, mostly because the model selects a wrong skill or none, and \citet{liu2026skillsrealistic} show that agents often fail to load the skills a task needs once they must select them.
Studies of tool selection find strong position and description biases when several tools can serve a request~\cite{blankenstein2026biasbusters,faghih2025toolpreferences}, but they report which tool is chosen rather than what the user loses.
Large-scale studies that collect skills from marketplaces and GitHub examine their security and reusability~\cite{liu2026skillswild,zhang2026reusable}.
To the best of our knowledge, no prior work has systematically studied what such a conflict does to the core functions of the installed skill, how the platform and the model resolve the conflict, whether the user is told, or at what point a conflict is decided and whether it can be stopped there.

Dependency conflicts between packages are usually declared or resolved by rules, such as version constraints, that a tool can check from package metadata before anything runs~\cite{wang2020watchman,vouillon2013coinstall}.
A conflict between skills is usually resolved by the model during the run instead, which makes these questions hard to answer in three ways.
\cone \emph{A conflict hides behind success.}
Two skills compete because they do the same job, and for the same reason either one can complete the task, so the likelier a conflict, the less it shows.
Its only trace is missing behavior specific to the installed skill, which is written in prose, mixed with examples and generic advice, and different for every skill.
\ctwo \emph{A conflict is resolved out of sight.}
Unless a personal skill overrides a project skill of the same name, the model chooses from a few hundred characters of description, without having to tell the user.
It sometimes reads the files of a skill instead of invoking it, and its choice can change from one run to the next.
What it sees depends on listing order, scope, and the model, which change together in practice.
\cthree \emph{A conflict is decided at run time, yet it must be caught before the similar skill does the work.}
The model chooses from the user's request and a few hundred characters per skill, so whether two skills conflict may depend on a request that does not exist when they are installed, and similar wording does not imply the same job.
Learning which pairs are risky requires the pairs that users actually combine, yet the personal skills and plugins that a user installs are rarely visible in public.

To tackle these challenges, we conduct a controlled empirical study of \cc on pairs of an installed skill \skillA and a similar skill \skillB.
For \cone, we derive checks for each skill from its text alone.
These checks are its core functions, the requirements that a competent model would not meet without the skill, quoted from its text, and checkable in the output.
The first three that apply to a task, in rank order, are fixed before any run, and the share of them fulfilled in a run is the \emph{fidelity}, which we report next to task completion.
To isolate what only \skillA provides, we also label each core function of \skillA as \emph{exclusive} when \skillB does not ask for it (\S\ref{sec:design-core}, \S\ref{sec:design-cases}).
For \ctwo, we observe the choice in the full record of tool calls, including reads of skill files.
This record separates \emph{substitution}, in which \skillB is used instead of \skillA, from \emph{interference}, in which \skillA is used but the result still changes.
We then change one factor at a time, running each pair with only \skillA, only \skillB, \skillA and \skillB, and \skillA with an unrelated skill, then with the listing order reversed and with \skillB moved to the personal or the plugin scope, on three models (\nRunsTotal runs).
Paired comparisons over hundreds of pairs separate real differences from run-to-run variation (\S\ref{sec:design-runs}, \S\ref{sec:design-analysis}).
For \cthree, we reconstruct from complete snapshots the skills that \cc would load in \nCorpusRepos repositories, pair each installed skill with similar skills in the same installation list and in other projects, and let a large language model (LLM) judge whether two skills do the same job.
A stratified sample of \nSampleThree judged pairs gives prevalence estimates and \nPairs experimental pairs of normative, capability, and script-bearing skills (\S\ref{sec:design-install}--\S\ref{sec:design-cases}).
Because a conflict is decided at run time, we then locate the first read of each skill in the record of tool calls, compare runs of the same pair and model that open different skills first, and test a guard at that read in a new experiment on the \nGuardCombos pair--model combinations at risk (\S\ref{sec:design-analysis}).

We organize the study around three research questions (RQs).
\begin{itemize}[leftmargin=*,nosep]
  \item \textbf{RQ1 (Impact).} \emph{When two similar skills are co-installed, do task completion and the core functions of the installed skill decline?}
  We also ask whether a decline comes from substitution or from interference, and how it differs across normative, capability, and script-bearing pairs.
  \item \textbf{RQ2 (Selection).} \emph{When two similar skills are co-installed, what determines which one is used, and does the model tell the user?}
  We examine listing order, scope, and the model.
  \item \textbf{RQ3 (Decision point).} \emph{At what point in a run is a conflict decided, and can a guard at that point keep the core functions of the installed skill?}
  We locate the first read of each skill in the record of tool calls and test a pre-tool hook at that read.
\end{itemize}

Our study makes the following major findings.
\textbf{\textit{A similar skill takes one in five runs away from the installed skill, yet task completion does not drop.}}
The use of the installed skill falls by \dUseA percentage points, against \dUseAC points with an unrelated skill, so a benchmark that checks only task completion would call co-installation harmless.
Its fidelity falls by \dFid points, or \dFidUniqueAll points on its exclusive core functions, and most of the loss comes from runs that do not use it.
\textbf{\textit{Where a skill is installed decides which one runs, and the final reply rarely says so.}}
A personal skill of the same name overrides the installed skill in every listing, yet models still read the overridden files in \pctPersonalReadA of runs.
A plugin skill is used in only \pctPluginB of runs, and reversing the listing order, which biases tool selection, changes the use of the installed skill by only \dOrder points.
When the similar skill replaces the installed one, the final reply names it in \pctNameUsedSubst of runs.
\textbf{\textit{A conflict is decided at the first read of a skill, almost always before the agent changes anything, and a guard at that read keeps the core functions of the installed skill.}}
Within the same pair and model, opening the similar skill first lowers fidelity on the exclusive core functions of the installed skill by \dFlipEx points.
In \pctNoChange of the runs that open the similar skill first, the agent has not yet changed any file.
In a new experiment, a pre-tool hook that denies the first read of the similar skill made the model switch to the installed skill in \pctGuardComply of cases and brought this fidelity back to the level of runs that open the installed skill first, with a significant gain for every model.

This paper makes the following contributions.
\begin{itemize}[leftmargin=*,nosep]
  \item \textbf{Skill conflicts at scale.} To the best of our knowledge, this is the first empirical study of conflicts between benign, co-installed skills that do the same job, and it shows from repository snapshots that such pairs are common and mostly involve normative skills.
  \item \textbf{Requirement-level evaluation.} We are the first to measure, requirement by requirement, what an installed skill loses to a similar skill, and to tell from the record of tool calls which skill ran as we vary listing order, scope, and the model.
  \item \textbf{Decision point and guard.} We are the first to locate the point at which a conflict is decided and to show that a pre-tool hook at that point keeps the core functions of the installed skill.
  \item \textbf{Implications and artifacts.} We derive implications for platforms, skill authors, and benchmark designers, and we release all cases, the harness, and all run logs.
\end{itemize}

\section{Background and Motivation}
\label{sec:background}

\subsection{Agent Skills in Claude Code}
\label{sec:bg-skills}

A skill is a directory whose entry point is a \skillmd file~\cite{claudecode2026skills,agentskills2026spec}.
Its YAML frontmatter holds a \texttt{description} of what the skill does and when to use it, and a \texttt{name}, which \cc makes optional, and its Markdown body holds the instructions the model follows once the skill runs.

\cc selects skills in two stages~\cite{claudecode2026skills}.
First, it places a listing of skill names and descriptions into the context of the model, truncating the combined \texttt{description} and \texttt{when\_to\_use} text of each skill at 1,536 characters.
Second, the model invokes a skill when the user's request matches its description, and only then is the body loaded.
The choice of the model between two similar skills therefore rests on a few hundred characters of text per skill, and the body, where the requirements of the skill usually live, is read only after the choice is made.

Skills can be installed at several scopes, summarized for \cc in Table~\ref{tab:scopes}, and plugins can be enabled for a user or for a whole repository~\cite{claudecode2026plugininstall}.
When two skills have the same name, the scope decides whether one overrides the other or both are listed~\cite{claudecode2026skills}.
We use \cc as the example, but Codex, Gemini CLI, and GitHub Copilot also load project skills and personal skills~\cite{codex2026skills,geminicli2026skills,copilot2026aboutskills}.
They differ on which skill wins when two have the same name.
Gemini CLI and the Copilot CLI prefer the project skill~\cite{geminicli2026skills,copilotcli2026config}, \cc prefers the personal skill, and Codex lists both~\cite{codex2026skills}, although the implementation guide of the format calls the project-first rule the universal convention~\cite{agentskills2026clientguide}.

\begin{table}[t]
  \caption{Skill scopes in \cc, our example platform, and what happens when two skills have the same name~\cite{claudecode2026skills}.}
  \label{tab:scopes}
  \small
  \begin{tabular}{@{}lll@{}}
    \toprule
    Scope & Location & Same name as a skill at another scope \\
    \midrule
    Enterprise & managed settings directory & overrides personal and project \\
    Personal & \texttt{\~{}/.claude/skills/<name>/} & overrides project \\
    Project & \texttt{.claude/skills/<name>/} (start dir.\ and parents) & overridden by personal and enterprise \\
    Nested & \texttt{<subdir>/.claude/skills/<name>/} & both load, directory-qualified \\
    Plugin & \texttt{<plugin>/skills/<name>/} & both load, namespaced \texttt{plugin:name} \\
    \bottomrule
  \end{tabular}
\end{table}

\subsection{Skill Conflicts}
\label{sec:bg-substitution}

The \emph{installed skill} \skillA is a skill that a project or user installed for a purpose.
A \emph{similar skill} \skillB is another skill that does the same job.
The two are \emph{co-installed} when they are visible in the same session.
The \emph{core functions} of a skill are the requirements that a competent model would not meet without it, that are quoted from its text, and that can be checked in the output of a given task (\S\ref{sec:design-core}).
The \emph{primary core functions} of \skillA are the first three that apply to a task, and the \emph{fidelity} of a run is the share of them fulfilled.

Two skills \emph{conflict} when co-installing them changes what the user gets from \skillA, in task completion or in fidelity, compared with installing \skillA alone.
A conflict can arise in two ways.
In \emph{substitution}, \skillA is not used because the platform or the model uses \skillB instead.
In \emph{interference}, \skillA is used, yet the result still changes, for example because instructions or scripts of \skillB change what it does.
A run may also use neither skill, which we report as a third outcome.
Throughout, we take the view of a user who expects the behavior of \skillA, and each request asks for its job.
Whatever the mechanism, the model may \emph{disclose} to the user that two similar skills were available or which one it used.

\begin{figure}[t]
  \centering
  \includegraphics[width=0.95\linewidth]{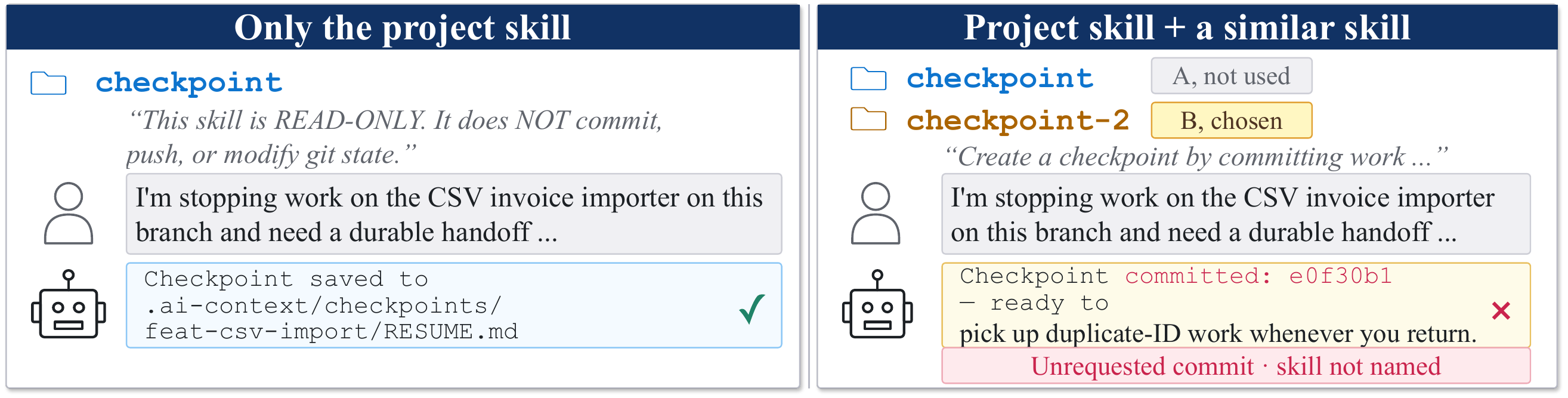}
  \caption{Two runs of one request with Haiku 4.5, abridged, with the project skill \skillA in \textcolor[HTML]{0D74CE}{blue}, the similar skill \skillB in \textcolor[HTML]{AB6400}{gold}, and the failure in \textcolor[HTML]{CA244D}{red}. With \skillB added under the directory \texttt{checkpoint-2}, the model uses it, commits the work that \skillA forbids, and does not say which skill it used.}
  \label{fig:motivating}
\end{figure}

Figure~\ref{fig:motivating} shows a clear case from our study.
The project skill \texttt{checkpoint} forbids any change to the git state, and with only this skill installed, the model saves a handoff note without committing.
With a similar skill of the same name added to the project as \texttt{checkpoint-2}, the model uses the similar skill instead, commits the work, and names no skill in its reply.
Both runs complete the task, so a check of task completion passes both, whereas the core function of the project skill fails in the second (challenge \cone).

\subsection{Skill Types}
\label{sec:bg-types}

The measure that can reveal a loss depends on what a skill contributes, so we distinguish three types of skills.

\emph{Normative skills} encode conventions, such as pull-request and commit-message formats or a coding style, which the documentation calls reference content~\cite{claudecode2026skills}.
A capable model often completes a task without any skill~\cite{han2026sweskillsbench}, so losing a normative skill leaves task completion unchanged and shows only in fidelity.

\emph{Capability skills} supply specific knowledge or tools without which the task is done poorly, as in SkillsBench, where curated skills raise the average pass rate from 33.9\% to 50.5\%~\cite{li2026skillsbench}.
Substituting such a skill can lower task completion as well as fidelity.

\emph{Script-bearing skills} ship executable scripts that their instructions run.
When such a skill is substituted, its scripts do not run at all, and when both skills are active, the scripts of one skill may run under the instructions of the other.
We therefore treat pairs whose installed skill bundles scripts as a type of their own.

\section{Study Design}
\label{sec:design}

\subsection{Overview}
\label{sec:design-overview}

\begin{figure}[t]
  \centering
  \includegraphics[width=0.92\linewidth]{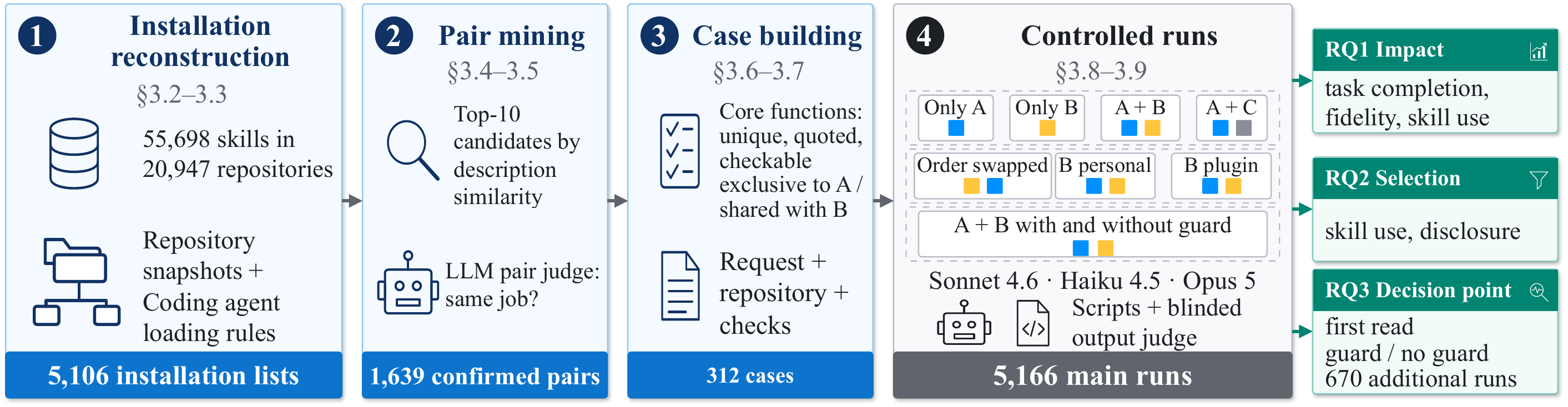}
  \caption{Overview of the study, with the installed skill \skillA in \textcolor[HTML]{0D74CE}{blue}, a similar skill \skillB in \textcolor[HTML]{AB6400}{gold}, and an unrelated skill \skillC in \textcolor[HTML]{8B8D98}{gray}.}
  \label{fig:overview}
\end{figure}

We designed the study to answer the three RQs with pairs that are or could plausibly be co-installed and a loss measure fixed in advance (Figure~\ref{fig:overview}).
We study \cc because Anthropic introduced skills for it before opening the format as a standard~\cite{anthropic2025skills}, it is the largest identifiable target of public skills~\cite{zhang2026reusable,galster2026configs}, and its documentation states how skills are loaded and overridden~\cite{claudecode2026skills}.
Because the full design costs about \costAll at API prices on one platform (\S\ref{sec:design-runs}), we run it on \cc only and check its main effects on Codex.
There, with \texttt{gpt-5.5}, \texttt{gpt-5.6-luna}, and \texttt{gpt-5.6-sol} on \nCodexPairs pairs judged in the same way, the similar skill lowered the use of \skillA by \codexUseA percentage points and fidelity on its first three exclusive core functions by \codexExcl, close to the \codexClaudeUseA and \codexClaudeExcl points on \cc for the same pairs, without lowering task completion.
The study reconstructs installations (\S\ref{sec:design-install}--\ref{sec:design-filter}), mines pairs (\S\ref{sec:design-pairs}--\ref{sec:design-sampling}), builds cases around the core functions of the installed skill (\S\ref{sec:design-core}--\ref{sec:design-cases}), and scores controlled runs (\S\ref{sec:design-runs}--\ref{sec:design-analysis}).
Sampling and rating follow published guidelines~\cite{baltes2022sampling,sigsoft2026empirical}, and two human raters check every LLM-assisted step on random samples (\S\ref{sec:design-human}).

\subsection{Installation Reconstruction}
\label{sec:design-install}

Learning which pairs are risky requires the pairs that users actually combine (challenge \cthree), so we first reconstruct which skills are co-installed.
We start from a corpus of public \skillmd files collected on \corpusSnapshot, with \nCorpusSkills distinct skills and \nCorpusCopies non-fork copies in \nCorpusRepos repositories.
The corpus was built mainly with GitHub code search, which indexes only default branches~\cite{ghcodesearch2026} and returns at most 1,000 results per query~\cite{ghcodesearch2026}.
To recover every skill, we therefore fetch the complete file tree of each repository at the last commit no later than its recorded push time, and we store the commit hash and the git object id of every file.
This pins the data, which repositories can otherwise rewrite or delete~\cite{gonzalezbarahona2012reproducibility,kalliamvakou2016promises}.

A skill counts as \emph{installed} when \cc would load it in a session (Table~\ref{tab:scopes})~\cite{claudecode2026skills,claudecode2026plugins}.
This covers skills in a project or nested \texttt{.claude/skills} directory of a real project, including resolved symbolic links, and skills in the personal directory kept in a dotfiles repository~\cite{zhu2025dotfiles}.
It also covers plugins placed under \texttt{.claude/skills} or enabled in the committed settings of the project, the two ways to share a plugin through a repository, but not other plugins~\cite{claudecode2026plugins}.
We exclude deeper nesting, which a maintainer states is unsupported~\cite{claudecodeissue18192}, the directories of other agents, documentation, example, and test paths~\cite{herbold2022tangling}, and vendored paths~\cite{linguist2026vendor}.

Because only some repositories are actually used with their skills, an LLM, GPT-6 Astra~\cite{openai2026gpt6astra}, assigns a type to each of the \nTypedRepos repositories with a skill in a \texttt{.claude/skills} directory or in a personal directory kept in dotfiles, based on where its skills sit, what else it contains, its README, and any plugin manifest, much as prior work identifies engineered software projects~\cite{munaiah2017reaper}.
Real projects (\nTypedReal) and dotfiles repositories (\nTypedDotfiles) contribute installation lists, whereas collections curated for others to pick from~\cite{wu2017curation} (\nTypedCollection) and single-skill repositories (\nTypedSingle) supply similar skills.
The remaining \nTypedUnsure stay undecided.

An \emph{installation list} is the set of installed skills that a session in one project or dotfiles repository sees.
We obtain \nLists installation lists, \nListsMulti of them with at least two distinct skills (Figure~\ref{fig:overview}).
The \emph{A pool} holds the \nApool distinct skills installed in real projects or dotfiles repositories, more than the corpus holds, because the snapshots include skills that code search missed.
The \emph{B pool} holds the \nBpool skills installed in another real project or listed in a plugin marketplace, which a user could plausibly add from elsewhere.

\subsection{Skill Filtering}
\label{sec:design-filter}

To keep only skills that the model can actually use, we downloaded by git object id the \nDownloaded distinct \skillmd files in the project, personal, and plugin skill directories of the snapshots.
A skill is kept only if \cc reads its frontmatter and the model can invoke it, which requires frontmatter on the first line~\cite{claudecode2026skills}, valid YAML~\cite{agentskills2026spec}, and no \texttt{disable-model-invocation} flag~\cite{claudecode2026skills}.
As in prior skill corpora, it must also have a non-empty name and description~\cite{liu2026skillsrealistic} and at least ten lines of instructions~\cite{liu2026skillswild}.
These checks keep \nValid files (\pctValid).
Versions of one skill count as one skill, so we group near-duplicates~\cite{lopes2017dejavu} into \emph{families} with MinHash over word 5-grams at a Jaccard threshold of 0.8~\cite{broder1997resemblance,lee2022dedup}, never pair two skills of one family, and use one installed skill per family (\nApoolFamilies families in the A pool).

\subsection{Candidate Retrieval and Confirmation}
\label{sec:design-pairs}

We pair skills in two ways because each captures a different way in which two skills become co-installed.
\emph{Within-project pairing} combines two skills in the same installation list, so it shows co-installation that actually exists.
\emph{Cross-project pairing} combines an installed skill \skillA with a skill \skillB from the B pool by a different author, which models a user who adds a skill from elsewhere, including the personal and plugin scopes that repository scans cannot observe.

To find candidates cheaply, we retrieve them by the text the model sees when choosing, namely the name and the combined description and \texttt{when\_to\_use} text (\S\ref{sec:bg-skills}).
We embed this text with Qwen3-Embedding-0.6B~\cite{zhang2025qwen3embedding} and call the cosine similarity of two embeddings their \emph{description similarity}.
For each installed skill, we keep the ten candidates with the highest description similarity, in the B pool or in the same installation list, excluding the same family and, for cross-project pairing, the same author or repository, as in retrieve-then-confirm pipelines for overlapping tools and skills~\cite{blankenstein2026biasbusters,liu2026toolscope,kang2026skillret}.
This yields \nCandCross cross-project and \nCandWithin within-project candidate pairs.

GPT-6 Astra, as the \emph{pair judge}, then decides whether two skills do the same job.
It answers whether, from names and descriptions alone, a request meant for \skillA could be routed to \skillB, and whether, after reading their bodies, the two do the same job, are near-duplicates, are related but different, or are unrelated.
A pair is \emph{confirmed} when the second answer is ``same job'' or ``near-duplicate'', and near-duplicates here are skills of different families (\S\ref{sec:design-filter}).

Before sampling, we calibrated a floor on description similarity with \nCalibPairs pairs, each judged in both orders because LLM judges are sensitive to position~\cite{zheng2023llmjudge}.
Following the recall target used to evaluate screening for systematic reviews~\cite{cohen2006wss,omaraeves2015textmining}, we take the highest floor that keeps at least 95\% of the confirmed pairs.
This floor is \floorCross for cross-project pairing (recall \recallCross) and \floorWithin for within-project pairing.
Confirmed pairs occur even at low description similarity, and below 0.9 only 10\% to 37\% of the cross-project pairs in each band are confirmed, so similar wording does not imply the same job and description similarity cannot stand in for the pair judge (challenge \cthree).
Each sampling unit is one installed skill per family, paired with its most similar candidate above the floor, which gives frames of \nFrameCross cross-project and \nFrameWithin within-project units.

\subsection{Sampling and Prevalence}
\label{sec:design-sampling}

We draw one stratified random sample to estimate prevalence and to select the experimental pairs.
The strata cross pairing, whether \skillA bundles scripts, and description-similarity band, and each pairing-by-script group is sized for a 95\% confidence level~\cite{kalliamvakou2014promises,un2008survey} and a $\pm$3\% margin.
Drawing units in a seeded random order yields \nSampleThree pairs, of which the pair judge confirms \nConfirmed (\nConfirmedCross of \nSampleThreeCross cross-project and \nConfirmedWithin of \nSampleThreeWithin within-project pairs).
Re-judging 200 random pairs reproduces the verdict almost perfectly (Cohen's $\kappa$ of \kappaJudgeRepeat), and two human raters validate the pair judge on \hvN random pairs (\S\ref{sec:design-human}).

To compare skill types, GPT-6 Astra labels the \nLabeledSkills skills in confirmed pairs as normative or capability by asking the counterfactual question of whether a capable model could complete the typical task of the skill without it.
Only specific, non-obvious knowledge or tools count as capability, which gives \nLabeledNormative normative and \nLabeledCapability capability skills (repeat-run $\kappa=$\kappaTypeRepeat, human validation in \S\ref{sec:design-human}).
A skill is script-bearing when its directory contains scripts that its \skillmd refers to~\cite{agentskills2026spec}.

\subsection{Core Function Extraction}
\label{sec:design-core}

Because the only trace of a conflict is missing behavior specific to the installed skill (challenge \cone), we extract the core functions of every skill in a confirmed pair before any run.
The extractor, GPT-6 Astra, sees one skill at a time, never its partner or any run, and lists the requirements of the skill that pass three tests.
A core function is \emph{unique}, meaning a competent model doing the same kind of task without the skill would usually not meet it, \emph{quoted} verbatim from the skill, and \emph{checkable} in written files, pull requests, commit messages, executed commands, or the final reply.
Examples, sample outputs, reference tables, and defaults that a tool adds anyway do not count, and each core function states one requirement, as in checklist-based evaluation~\cite{qin2024infobench,cook2024tick}, with prohibitions counted as requirements.

Core functions are ranked by the strength of the wording (an explicit MUST or NEVER over SHOULD over plain instructions), closeness to the stated purpose of the skill, whether the effect lands in a deliverable or concerns safety, and order of appearance, and up to ten are kept.
Across \nLabeledSkills skills, the extractor lists \nCoreItems core functions, \meanCoreItems per skill on average.
The primary core functions are the first three, in rank order, that apply to the request of the case (\S\ref{sec:design-cases}), and a conditional core function is scored only in runs where its condition occurs.

In a second pass, each core function of \skillA is compared with the full \skillmd and the core functions of \skillB in its pair, again without any run, and labeled \emph{exclusive} to \skillA when \skillB does not ask for the same specific behavior and \emph{shared} otherwise.
Of the \nCoreApplicableA applicable core functions of the experimental pairs, \pctUniqueItems are exclusive, and two human raters repeat both steps independently on \hvCorePairs random pairs (\S\ref{sec:design-human}).

\subsection{Pair Selection and Case Construction}
\label{sec:design-cases}

A pair enters the experiment only if a loss can be measured for it.
The pair must be confirmed, \skillA must run fully in an offline container and have at least three core functions, at least one of which applies to the request, and the directories of the two skills must be retrievable at the snapshot commit.
Offline runnability, judged by GPT-6 Astra from the text of the skill, holds for \nRunnableFull of the \nRunnableCandidates installed skills in confirmed pairs (\pctRunnableFull, human check in \S\ref{sec:design-human}).
The experimental pairs are all eligible pairs in a nested sample sized for $\pm$5\% (\nSampleFive units), plus all eligible pairs of the two rare capability cells in the $\pm$3\% sample, weighted by inverse inclusion probability in population estimates.
Table~\ref{tab:cells} lists the \nPairs pairs in six cells, where the script-bearing cells are defined by \skillA bundling scripts, and each \skillA appears only once.
Of the cross-project pairs, \nScopePairs have the same name.

\begin{table}[t]
  \caption{Confirmed pairs, experimental pairs, and the analyzed runs and tool calls of the experimental pairs per cell.}
  \label{tab:cells}
  \small
  \begin{tabular}{@{}llrrrr@{}}
    \toprule
    Pairing & Skill type & Confirmed pairs & Experimental pairs & Runs & Tool calls \\
    \midrule
    \multirow{3}{*}{Cross-project} & Normative & 438 & 98 & 1,638 & 34,216 \\
     & Capability & 223 & 41 & 723 & 23,223 \\
     & Script-bearing & 542 & 89 & 1,541 & 43,737 \\
    \midrule
    \multirow{3}{*}{Within-project} & Normative & 150 & 34 & 510 & 11,194 \\
     & Capability & 95 & 14 & 210 & 8,415 \\
     & Script-bearing & 191 & 36 & 540 & 16,294 \\
    \midrule
    Total & & 1,639 & 312 & 5,162 & 137,079 \\
    \bottomrule
  \end{tabular}
\end{table}

\begin{figure}[t]
  \centering
  \includegraphics[width=0.9\linewidth]{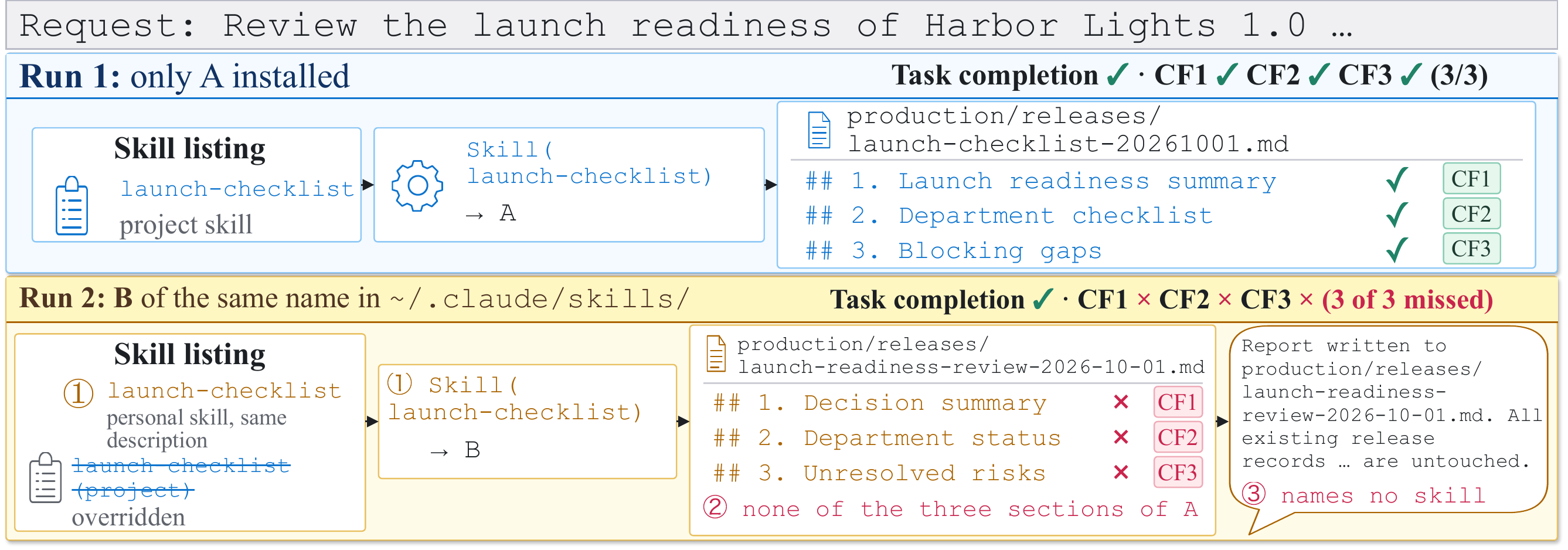}
  \caption{A case and two of its runs with Opus 5, abridged, with \skillA in \textcolor[HTML]{0D74CE}{blue}, \skillB in \textcolor[HTML]{AB6400}{gold}, and missed core functions in \textcolor[HTML]{CA244D}{red}, where CF1--CF3 are the three primary core functions of \skillA. With \skillB of the same name in the personal directory, \cc runs \skillB (\ding{172}), the task is still completed but none of CF1--CF3 is fulfilled (\ding{173}), and the final reply names no skill (\ding{174}).}
  \label{fig:case}
\end{figure}

For every pair we build one case that all configurations share.
GPT-6 Astra drafts a request for the job of \skillA, a starting repository of at most 15 files, the applicability of each core function of \skillA, and a check for each applicable core function and for task completion.
The request states the goal and the facts a developer would mention, but never the name of a skill, its conventions, or phrases from its description, and the repository contains no instructions addressed to an agent.
A check is either a script over the final repository, the executed commands, and the final reply, which returns fulfilled, not fulfilled, or undecidable, or a yes/no question for the output judge, an LLM described in \S\ref{sec:design-analysis}.
Task completion is judged on substance, so any competent way of doing the work counts.
Every script is also run on the untouched starting repository, and core functions whose script already passes there are flagged for a sensitivity analysis.
In the case of Figure~\ref{fig:case}, both runs complete the task, so only the three primary core functions reveal that the second run followed \skillB instead of \skillA.

\subsection{Configurations and Setup}
\label{sec:design-runs}

Because a conflict is resolved out of sight (challenge \ctwo), we change one factor at a time.
RQ1 compares the only-\skillA, the only-\skillB, and the A+B configuration on every pair, with the same request and starting repository (Table~\ref{tab:configs}).
The only-\skillB configuration shows what \skillB delivers on its own, and paired comparisons over hundreds of pairs separate real differences from run-to-run variation.
To separate the effect of a similar skill from that of one more skill, the A+C configuration installs an unrelated skill \skillC instead of \skillB on all \nPairs pairs.
\skillC is one of \nCskills skills of other cases whose description similarity to \skillA and to \skillB is at most 0.435, the median over 200,000 random pairs, and which sits on the same side of \skillA in the listing as \skillB and has a description of comparable length.
RQ2 adds a swapped configuration, which reverses the listing order of the two skills.
Because the scope rules in Table~\ref{tab:scopes} apply to skills of the same name, RQ2 also adds two scope configurations for the \nScopePairs pairs whose skills have the same name, all from cross-project pairing.
In the personal-scope configuration, \skillB sits in the personal directory and overrides \skillA, and in the plugin-scope configuration, \skillB sits in a plugin, the two skills are listed, and the model chooses.

\begin{table}[t]
  \caption{Configurations with their analyzed runs, tool calls, and hours of agent time, each run once per pair and model, followed by the guard experiment and the repeated runs.}
  \label{tab:configs}
  \small
  \setlength{\tabcolsep}{4pt}
  \begin{tabular}{@{}lllrrrrl@{}}
    \toprule
    Configuration & \skillA & \skillB & Pairs & Runs & Tool calls & Hours & RQ \\
    \midrule
    Only \skillA & project & -- & 312 & 935 & 24,617 & 81.3 & 1, 3 \\
    Only \skillB & -- & project & 312 & 936 & 24,400 & 78.7 & 1 \\
    A+B & project & project & 312 & 934 & 24,555 & 81.6 & 1--3 \\
    Swapped & project & project, renamed & 312 & 936 & 24,633 & 81.9 & 2, 3 \\
    Personal scope & project & personal & 81 & 243 & 6,891 & 22.3 & 2 \\
    Plugin scope & project & plugin & 81 & 243 & 6,749 & 22.8 & 2 \\
    A+C & project & \skillC in project & 312 & 935 & 25,234 & 68.8 & 1 \\
    \midrule
    A+B with and without guard & project & project & 119 & 670 & 18,192 & 54.0 & 3 \\
    Repeats beyond the first run & & & 224 & 536 & 14,023 & 51.1 & -- \\
    \midrule
    Total & & & & 6,368 & 169,294 & 542.4 & \\
    \bottomrule
  \end{tabular}
\end{table}

We measured how \cc 2.1.283 lists skills to the model.
The listing groups skills by source (personal, project, plugin, bundled) and sorts each group by directory name, which is the displayed name, so two skills have the \emph{same name} when their directories do.
A personal skill overrides a project skill of the same name, which then disappears from the listing, whereas a plugin skill of the same name is listed with a plugin prefix next to the project skill.
Accordingly, \skillB receives the suffix \texttt{-2} when it has the same name as \skillA in the same project, and the swapped configuration prefixes the directory name of \skillB with one character that reverses the order.
The whole listing, which also contains the bundled skills, is capped at 1\% of the context window~\cite{claudecode2026skills}.
We measured this cap as 8,000 characters for Sonnet 4.6 and Haiku 4.5 and 6,000 for Opus 5.
Under it, \pctCapDropOpusAny of the listings shown to Opus 5 in the A+B configuration would lose at least one of the two descriptions.
We therefore raise the cap to 20,000 characters for all models and record in every run the listing the model received.

Each run starts in a fresh container with no host directory mounted and runs \cc 2.1.283 non-interactively for at most 80 turns and 25 minutes.
All experiments use three Claude models from the Sonnet, Haiku, and Opus tiers~\cite{anthropic2026models}, \texttt{claude-sonnet-4-6} (Sonnet 4.6), \texttt{claude-haiku-4-5-\allowbreak 20251001} (Haiku 4.5), and \texttt{claude-opus-5} (Opus 5), with one run per model, configuration, and pair in a seeded random order.
Timeouts are scored as they are, and only infrastructure failures are re-run.
The container reaches only the model application programming interface (API) and the PyPI and npm registries, and credentials never enter it.
In total, we analyze \nRunsUsable of the \nRunsTotal runs of the configurations, \nRunsTimeout of which timed out.
With the guard experiment (\S\ref{sec:design-analysis}) and the repeated runs below, the study comprises \nRunsAll runs, \hoursAll hours of agent time, \toolCallsAll tool calls, \tokensInAll input and \tokensOutAll output tokens, and about \costAll at the API prices that \cc reports (Table~\ref{tab:configs}).

Each combination of model, configuration, and pair runs once, a single-sample estimate of pass@1~\cite{chen2021evaluatinglargelanguagemodels}, because three runs give nearly the same results at three times the cost.
In the experimental pairs, \nRepeatCombos combinations, drawn by the seeded random run order, were run two or three times (\nRepeatCombosRuns runs).
Whether \skillA is used agrees between their runs in \pctRepUseAgree of pairwise comparisons (intraclass correlation \iccRepUse), and whether its core functions are fulfilled, as decided by scripts, has an intraclass correlation of \iccRepFulfilAll.
Every analysis therefore uses the first completed run of each combination.

\subsection{Measures and Analysis}
\label{sec:design-analysis}

For every run we record the fidelity, task completion, skill use, and, in configurations that co-install \skillA and \skillB, disclosure.
\emph{Skill use} means that the model invokes a skill through the Skill tool or reads or runs files in its installation directory, because models often read the files of a skill directly instead of invoking it (challenge \ctwo), and we record Skill-tool invocations separately.
\emph{Disclosure} means that the final reply says that more than one suitable skill exists, says which one was used, or asks the user to choose.
Scripts decide \nTopItemsScript of the \nTopItems primary core functions inside the offline container, and the output judge decides the other \nTopItemsByJudge, any script check that returns undecidable, and all but \nCompletionProgram checks of task completion.
The \emph{output judge} is GPT-6 Astra, which sees the final repository, the full record of tool calls with skill names and paths masked, and the final reply, but neither the configuration nor which skill is \skillA, because LLM judges are sensitive to cues unrelated to the answer~\cite{zheng2023llmjudge}.
A conditional core function whose condition did not occur counts as not applicable.

For RQ1, we compare the A+B configuration with the only-\skillA configuration and report paired risk differences in fidelity, task completion, and the use of \skillA, per model and pooled.
Each difference is taken within pair and model, the pooled value averages the three models within each pair, and estimates are unweighted averages over the experimental pairs with 95\% confidence intervals (CIs) computed over pairs, with tests Holm-corrected over six cells and two outcomes, fidelity and task completion~\cite{holm1979sequential}.
Following \citet{song2026skillshadowing}, we condition on skill use, so a loss in a run that used \skillB but not \skillA is substitution, a loss in a run that used \skillA is interference, and runs that use neither skill form a third group.
As control checks, \skillA alone should mostly fulfill its core functions and \skillB alone should mostly not, and we report the loss separately on the exclusive and the shared core functions (\S\ref{sec:design-core}).
Sensitivity analyses use the first primary core function only, all applicable core functions, and only the core functions whose script does not pass on the starting repository.

For RQ2, the swapped configuration compares each pair with itself under the reversed listing order.
The scope configurations separate an override by the platform from a choice by the model, and model effects are estimated on all runs.

For RQ3 (challenge \cthree), the \emph{first read} of a skill in a run with \skillA and \skillB co-installed is the first tool call that invokes it through the Skill tool or reads or runs a file in its directory.
When the runs of a pair--model combination in the A+B and the swapped configuration open different skills first, the combination compares opening \skillB first with opening \skillA first within the pair.
We average over the two directions of this flip, which removes any effect of the configuration, and use a pair-level bootstrap with a Holm correction over three measures and four groups (pooled and per model).
The \emph{guard} is a pre-tool hook that denies the first read of \skillB while \skillA is unread and tells the model to use \skillA, without naming any core function, and denies \skillB once \skillA is in use.
A run \emph{reaches for} a skill first when its first attempt to read a skill targets that skill, whether or not the hook denies it.
The guard acts only when a run reaches for \skillB, so we rerun the \nGuardCombos pair--model combinations (\nGuardPairs pairs) in which the model opened \skillB first in the main experiment, each once with and once without the hook in random order.
The \nGuardCombosScript combinations whose outcome scripts decide run a second time.
All \nGuardRuns runs use the image, models, and settings of the main experiment.
We compare the two configurations on the fidelity of \skillA, judged as in the main experiment, over all runs and over the runs that reach for \skillB first in both configurations, and read compliance from the log of the hook.

\subsection{Use of LLMs and Human Validation}
\label{sec:design-human}
\label{sec:design-llm}

LLMs assisted repository typing (\S\ref{sec:design-install}), pair confirmation (\S\ref{sec:design-pairs}), skill labeling (\S\ref{sec:design-sampling}), core-function extraction and labeling (\S\ref{sec:design-core}), runnability judgments and case drafting (\S\ref{sec:design-cases}), and output judging (\S\ref{sec:design-analysis}), and every step was checked and guided by humans as described next.

Because LLM annotators and judges reach agreement close to that between humans on some tasks~\cite{ahmed2025llmannotation,zheng2023llmjudge} but not on others~\cite{ahmed2025llmannotation,zhuge2024agentjudge}, two human raters validate every LLM-assisted step on a random sample.
For each step, both raters label the same sample independently and without seeing the labels of the LLM, and they resolve disagreements by discussion~\cite{sigsoft2026empirical}.
All labeling, by the LLM and by the raters, is done without access to any run, except the output verdicts and their check, whose items are runs.
Samples have \hvN items for repository types, pair verdicts, skill types, and output verdicts, a size that related fields commonly use~\cite{baltes2026llmguidelines}, drawn at random or, for the output judge, stratified by verdict, and \hvSmallN skills for offline runnability, \hvSmallN cases for applicability, and \hvCorePairs pairs for core functions.
We report Krippendorff's $\alpha$, read against the thresholds of 0.8 for firm and 0.667 for tentative conclusions~\cite{baltes2026llmguidelines}.
Between the raters and between their consensus and the LLM, $\alpha$ is \hvRepoHH and \hvRepoLLM for repository types, \hvPairHH and \hvPairLLM for the pair judge, and \hvTypeHH and \hvTypeLLM for skill types.
For the output judge, it is \hvJudgeItemHH and \hvJudgeItemLLM on core functions, \hvJudgeCompHH and \hvJudgeCompLLM on task completion, and \hvJudgeDiscHH and \hvJudgeDiscLLM on disclosure.
For offline runnability, it is \hvRunHH and \hvRunLLM, and for whether each core function applies to the request of a case, it is \hvApplyHH and \hvApplyLLM.
For core functions, it is \hvCoreHH and \hvCoreLLM on which requirements of a skill are core functions, and \hvUniqueHH and \hvUniqueLLM on the exclusive-or-shared label.

\section{Results}
\label{sec:results}

\subsection{Prevalence}
\label{sec:prevalence}

Skills that do the same job are common (challenge \cthree, \S\ref{sec:design-sampling}).
For an estimated \pctSameJobCross of installed skills (95\% CI \ciSameJobCross), or about \nSameJobCross families, the most similar skill by another author does the same job, and for \pctSameJobWithin (95\% CI \ciSameJobWithin), or about \nSameJobWithin families, the most similar skill in the same installation list does, both conservative because the pair judge sees only the most similar candidate.
Copied collections bring many of these pairs.
In the \nBulkLists projects that copy a whole collection, with up to \maxBulkSkills skills, \pctBulkSameJob of the judged skills have a skill that does the same job in the same list, against \pctOtherSameJob in other lists.
Overall, in \pctWithinOutside of the confirmed within-project pairs at least one skill also exists outside the project, in a collection, a plugin, or another repository, and in the other \pctWithinOwn both skills exist only in that project.
Weighted to the frame, the installed skill is normative in \pctNormConfCross of the confirmed cross-project pairs and \pctNormConfWithin of the within-project ones, a capability skill in \pctCapConfCross and \pctCapConfWithin, and script-bearing in about \pctScriptConf.

\finding{Skills that can conflict are common. Nearly one in four installed skills already has a skill that does the same job in the same installation list, nearly two in three have one by another author, and \pctBulkSameJob do in projects that copy a whole skill collection.}
\finding{Most pairs that can conflict involve normative skills (\pctNormConfCross of cross-project and \pctNormConfWithin of within-project pairs), followed by capability skills (\pctCapConfCross and \pctCapConfWithin).}

\subsection{RQ1: Impact of Conflicts}
\label{sec:rq1}

RQ1 asks whether and why co-installing \skillB lowers task completion and the fidelity of \skillA, which we answer by comparing the A+B configuration with the only-\skillA configuration pair by pair (\S\ref{sec:design-analysis}).

\begin{figure}[t]
  \centering
  \includegraphics[width=\linewidth]{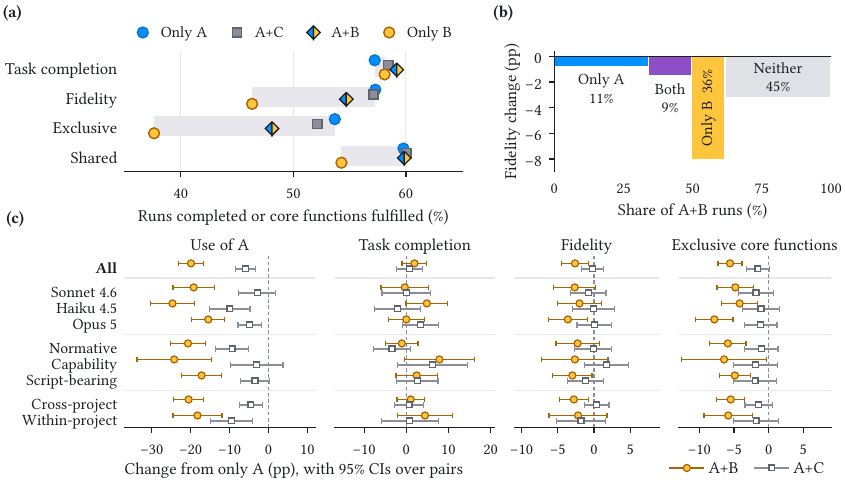}
  \caption{(a) Level of each configuration, with a gray band from only-\skillB to only-\skillA, (b) the A+B runs by the skills they use, where the area of a bar is its share of the loss of fidelity, and (c) the change from only-\skillA with 95\% CIs over pairs. Fidelity counts the primary core functions, and the exclusive and shared ones count all that apply.}
  \label{fig:rq1}
\end{figure}

\paragraph{Control checks}
The fidelity measure separates the two skills, whereas task completion does not (gray bands in Figure~\ref{fig:rq1}a).
With only \skillA installed, \fidOnlyA of its primary core functions are fulfilled, and with only \skillB installed, \fidOnlyB (\dFidB percentage points, pp, lower, 95\% CI \ciFidB), at equal task completion (57.2\% and 58.1\%).
The split into exclusive and shared core functions separates them further, as \skillB alone fulfills \dFidBUnique~pp fewer of the exclusive ones than \skillA alone, against \dFidBShared~pp fewer of the shared ones (\S\ref{sec:design-core}).

\paragraph{Task completion versus fidelity}
A similar skill takes runs away from the installed skill without failing the task, so the conflict hides behind success (challenge \cone, Figure~\ref{fig:rq1}).
Co-installing \skillB lowers the share of runs that use \skillA by \dUseA~pp (95\% CI \ciUseA), with every model (Figure~\ref{fig:rq1}c).
Task completion does not drop (\dTC~pp, 95\% CI \ciTC), whereas the fidelity of \skillA falls by \dFid~pp (\ciFid), a quarter of the \dFidB~pp that separate \skillB alone from \skillA alone (Figure~\ref{fig:rq1}a), with a lower point estimate for every model.
The loss concentrates on what only \skillA provides.
On the core functions exclusive to \skillA, fidelity falls by \dFidUniqueAll~pp (95\% CI \ciFidUniqueAll), significantly with every model, whereas on the shared ones it does not change (\dFidShared~pp, \ciFidShared).
A check of task completion therefore sees no harm, whereas the core functions that only \skillA provides are lost.

\paragraph{Similar versus unrelated skill}
The loss comes from a skill that does the same job, not from installing one more skill.
An unrelated skill lowers the use of \skillA by only \dUseAC~pp (\ciUseAC, gray in Figure~\ref{fig:rq1}c), is used in \pctCUsed of runs and instead of \skillA in \pctCInstead, and leaves its fidelity unchanged (\dFidAC~pp).
The similar skill lowers the use of \skillA by a further \dUseBothVsAC~pp (\ciUseBothVsAC) and its fidelity by \dFidBothVsAC~pp (95\% CI \ciFidBothVsAC).

\paragraph{Substitution versus interference}
Most of the loss occurs in runs that do not use \skillA at all (Figure~\ref{fig:rq1}b).
With \skillB co-installed, \pctBothOnlyB of runs use only \skillB and \pctBothNeither use neither skill, against \pctOnlyANoUse that do not use \skillA in the only-\skillA configuration (Figure~\ref{fig:rq2}).
A substituted run loses 8.0~pp of fidelity against the only-\skillA run, the deepest bar in Figure~\ref{fig:rq1}b.
Substituted runs carry \shareSubst of the fidelity loss and runs that use neither skill \shareNeither, whereas runs that use \skillA, where interference would show, carry only \shareInterf.
With \skillC, \pctNeitherAC of runs use neither skill and \pctCInstead use only \skillC, so beyond one more skill, \skillB adds mainly substitution.

\paragraph{Skill types and pairings}
The drop in the use of \skillA appears in every skill type and pairing, and so do a lower point estimate of fidelity and a loss on the exclusive core functions (Figure~\ref{fig:rq1}c).
Within projects, the use of \skillA drops by \dUseAWithin~pp, so the conflict also occurs for the pairs that projects actually co-install.

\paragraph{Sensitivity}
The loss grows with the number of core functions counted, from 1.7~pp with the first primary core function alone to \dFid~pp with the three primary ones and 4.1~pp with all applicable core functions (95\% CI [$-5.5$, $-2.6$]).
Without the core functions whose script already passes on the starting repository, it is still 2.3~pp ([$-4.1$, $-0.4$]).
Scripts confirm the loss.
It is larger on the core functions that scripts decide in both configurations (\dFidScript~pp, \ciFidScript) than on those that the output judge decides (\dFidJudge~pp).

\finding{A similar skill takes work away from the installed skill while task completion stays flat. Its use drops by \dUseA~pp and its fidelity by \dFid~pp, or \dFidUniqueAll~pp on the core functions exclusive to \skillA, whereas an unrelated skill lowers its use by only \dUseAC~pp. Most of the loss comes from runs that do not use \skillA.}

\subsection{RQ2: Selection and Disclosure}
\label{sec:rq2}

RQ2 asks how the choice between the two skills, made out of sight (challenge \ctwo), depends on listing order, scope, and the model, and whether the user is told.

\begin{figure}[t]
  \centering
  \includegraphics[width=\linewidth]{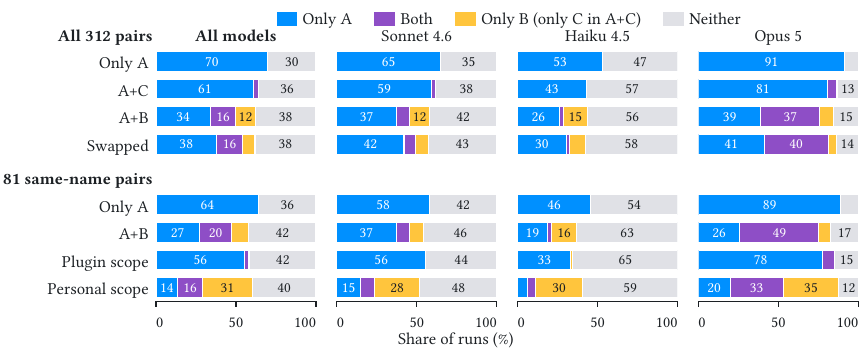}
  \caption{Skill use per configuration as the share of runs, over the three models and per model. The lower group covers the \nScopePairs same-name pairs with \skillB in the project (A+B), in a plugin, or in the personal directory, which removes \skillA from the listing.}
  \label{fig:rq2}
\end{figure}

\paragraph{Listing order}
Listing order barely affects which skill is used (Figure~\ref{fig:rq2}).
The swapped configuration reverses the order by renaming \skillB with a one-character prefix, so we separate the two effects by contrasting the pairs that listed \skillA first with the others.
Order changes the use of \skillA by only \dOrder~pp (95\% CI \ciOrder), whereas renaming \skillB adds \dRename~pp, in contrast to the position bias reported for selection among functionally equivalent tools~\cite{blankenstein2026biasbusters}.

\paragraph{Scope}
For skills of the same name, scope decides which skill is used (lower group of Figure~\ref{fig:rq2}).
With \skillB in the personal directory, \skillA appears in none of the 243 listings, as the listing rules predict, and the share of runs that use \skillA drops by \dUsePersonal~pp (\ciUsePersonal).
The model uses \skillB in \pctPersonalB of these runs, as in the second run of Figure~\ref{fig:case}, yet in \pctPersonalReadA it still reads the files of \skillA from the project directory, and the fidelity of \skillA is \dFidPersonal~pp below the only-\skillA configuration (95\% CI \ciFidPersonal).
With \skillB in a plugin, the model uses \skillA in \pctPluginA of runs and \skillB in only \pctPluginB, and the use of \skillA drops by 6.2~pp, against 16.9~pp when the same pairs have \skillB in the project.

\paragraph{Model}
The models differ sharply in every configuration (columns of Figure~\ref{fig:rq2}).
In the A+B configuration, Opus 5 uses both skills in 36.8\% of runs and neither in 15.5\%, whereas Haiku 4.5 uses neither in 56.4\%.
Opus 5 even reads the files of \skillA in \pctPersonalReadAopus of the personal-scope runs, where \skillA is not listed.

\paragraph{Disclosure}
A substitution is almost never disclosed.
Among the \nSubstRuns runs in which \skillB replaced \skillA, the final reply says that more than one suitable skill exists in 0\%, names the skill it used in \pctNameUsedSubst, and asks the user to choose in 0\%.
Over all A+B runs, the final reply names the skill it used in \pctNameUsedAB and says that more than one suitable skill exists in \pctMultipleAB, mostly with Opus 5 (\pctNameUsedABopus and \pctMultipleABopus), and it never asks the user to choose.

\finding{Where a skill is installed decides which one runs, and the final reply rarely says so. For skills of the same name, a personal \skillB overrides \skillA in every listing, yet models still read the files of \skillA in \pctPersonalReadA of runs. A plugin \skillB is used in only \pctPluginB of runs, and listing order changes the use of \skillA by only \dOrder~pp. When \skillB replaces \skillA, the final reply names the skill used in \pctNameUsedSubst of runs.}

\subsection{RQ3: Decision Point}
\label{sec:rq3}

RQ3 asks at what point in a run a conflict is decided and whether a guard at that point keeps the core functions of \skillA (challenge \cthree).

\begin{figure}[t]
  \centering
  \includegraphics[width=\linewidth]{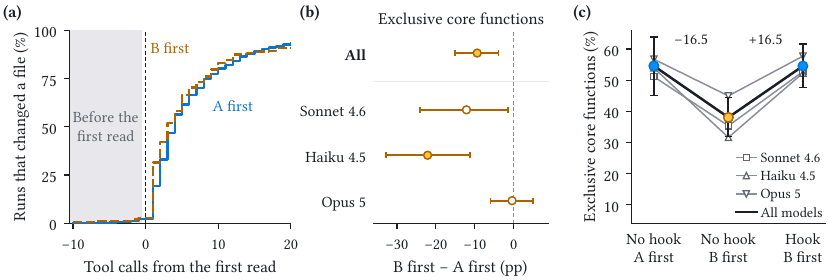}
  \caption{(a) Runs of the A+B and swapped configurations that have changed a file, by tool calls from the first read of a skill, (b) the change on the exclusive core functions of \skillA when the same pair and model open \skillB first instead of \skillA first, and (c) fidelity on these core functions in the guard experiment, by the skill that a run reached for first. Intervals are 95\% pair-bootstrap CIs, and hollow markers in (b) are not significant after Holm correction.}
  \label{fig:rq3}
\end{figure}

\paragraph{When it is decided}
A conflict is decided at the first read of a skill (Figure~\ref{fig:rq3}).
In \nFlipCombos pair--model combinations (\nFlipPairs pairs), the A+B and the swapped configuration, which differ only in the listing order and a one-character prefix on \skillB, open different skills first, and up to that read their runs look alike (median 6 against 6 tool calls).
There, the run that opens \skillB first fulfills \dFlipEx~pp fewer exclusive core functions of \skillA (95\% CI \ciFlipEx, Holm-adjusted $p = \pFlipEx$) and \dFlipAll~pp fewer applicable ones (\ciFlipAll).
It is \dFlipExHaiku~pp with Haiku 4.5, \dFlipExSonnet~pp with Sonnet 4.6, not significant after correction, and none with Opus 5, which often reads both skills (Figure~\ref{fig:rq3}b).
The first read comes before the work, as \pctNoChange of the runs that open \skillB first change no file before they open it, and files start to change only after the first read (Figure~\ref{fig:rq3}a).

\paragraph{What is lost}
When \skillB is opened first, what is lost is what only \skillA asks for.
Of the core functions fulfilled with only \skillA, \pctExLost of the exclusive ones are lost against \pctShLost of the shared ones, with the same split for every model, and each exclusive core function adds 1.9~pp of loss (95\% CI [0.9, 2.9]).

\paragraph{A guard at the first read}
The guard keeps the core functions of \skillA (Figure~\ref{fig:rq3}c).
When the hook denied a first reach for \skillB, the model switched to \skillA in \pctGuardComply of the \nGuardDenied cases (Sonnet 4.6 \pctGuardComplySonnet, Haiku 4.5 \pctGuardComplyHaiku, Opus 5 \pctGuardComplyOpus).
Over all runs, the hook raised fidelity on the exclusive core functions of \skillA by \guardITTex~pp (95\% CI \ciGuardITTex) and on all applicable core functions by \guardITTall~pp (\ciGuardITTall).
Both configurations reached for \skillB first about equally often (\pctGuardBFirstG of runs with the hook, \pctGuardBFirstC without), and where both runs of a combination did, the gain was \guardFiredPairedEx~pp (\ciGuardFiredPairedEx).
Among the runs that reached for \skillB first, the gain was significant with every model (\guardFiredExSonnet~pp with Sonnet 4.6, \guardFiredExHaiku~pp with Haiku 4.5, and \guardFiredExOpus~pp with Opus 5).
Runs redirected by the hook did not differ from runs without the hook that opened \skillA first ($\guardVsAEx$~pp, \ciGuardVsAEx), so the hook recovered \pctGuardRecoverEx of the \guardLossEx~pp that opening \skillB first costs without it, and task completion did not change ($+\guardCompITT$~pp, \ciGuardCompITT).
The first read is thus both the point where a conflict is decided and the point where it can be stopped.

\finding{A conflict is decided at the first read of a skill, which in \pctNoChange of runs comes before the agent changes anything, and what is lost is what only \skillA asks for. Opening \skillB first lowers fidelity on the exclusive core functions of \skillA by \dFlipEx~pp. A pre-tool hook at the first read of \skillB made the model switch to \skillA after \pctGuardComply of denials and brought fidelity on the exclusive core functions back to the level of runs that open \skillA first, with a significant gain for every model.}

\section{Discussion}
\label{sec:discussion}

\subsection{Implications for Skill Evaluation}
\label{sec:disc-metric}

A conflict hides behind success (challenge \cone), so skill evaluations should report what a skill uniquely contributes as well as whether the task is completed.
A benchmark that scores task completion alone would have rated co-installation as harmless in our runs, although the use of the installed skill fell by \dUseA percentage points and its fidelity by \dFidUniqueAll points on its exclusive core functions (\S\ref{sec:rq1}).
Core functions add little cost to a benchmark, because they are extracted once per skill, checked by scripts where possible, and frozen before any run, and benchmarks that already check the requirements of a skill~\cite{han2026skilluse,shaposhnikov2026framework} could co-install a similar skill with the one under test.

\subsection{Implications for Agent Platforms}
\label{sec:disc-platform}

A conflict is resolved out of sight (challenge \ctwo), but platforms can make substitution visible.
They can warn at session start when a personal skill overrides a project skill of the same name, as the issue report proposes~\cite{seasoned2026issue303}, and state which copy and scope a running skill came from.
Platforms can also act at the first read of a skill, where a conflict is decided (challenge \cthree, \S\ref{sec:rq3}).
There, a pre-tool hook can point the model to the installed skill, as our guard does, and the platform can pin the answer once per pair, as a lockfile does.
The final reply rarely fills this gap today (\S\ref{sec:rq2}), and showing which skill ran would also expose the runs that use neither skill, whose share rises from \pctOnlyANoUse to \pctBothNeither with a similar skill.

Finally, platforms can let a project declare which skill must win.
Under the current precedence~\cite{claudecode2026skills}, a team that ships a project skill cannot guarantee that its members run its version, and the override is neither visible nor complete (\S\ref{sec:rq2}).

\subsection{Implications for Skill Authors and Users}
\label{sec:disc-authors}

Authors control the choice mainly through the name and description, the only text the model sees when choosing.
A distinctive name avoids overrides by the platform, and a description that states the specific tasks of the skill reduces the chance that a request meant for another skill is routed to it.
Users with skills from several sources can audit their personal directory for overriding skills, and projects that copy a whole collection can remove the skills that duplicate ones they already install, since copied collections bring many pairs that can conflict (\S\ref{sec:prevalence}).

\section{Threats to Validity}
\label{sec:threats}

\paragraph{Internal validity}
Runs are nondeterministic, and each combination of model, configuration, and pair runs once.
The \nRepeatCombos combinations that we ran two or three times agree on the use of \skillA in \pctRepUseAgree of comparisons, with intraclass correlations of \iccRepUse for the use of \skillA and \iccRepFulfilAll for its core functions (\S\ref{sec:design-runs}).
Three runs would thus give nearly the same results at three times the cost, and every comparison is paired within pair and model.
Which skill a run opens first is observed rather than assigned, but the guard experiment randomizes the hook at the first read and shows that changing this read changes the result (\S\ref{sec:rq3}).

\paragraph{External validity}
We run the full design on \cc, the largest target of public skills, because it costs about \costAll on one platform.
A check on Codex with three OpenAI models on \nCodexPairs pairs finds effects of similar size, as the similar skill lowers the use of \skillA by \codexUseA percentage points against \codexClaudeUseA on \cc for the same pairs (\S\ref{sec:design-overview}), so the main effects are not specific to one platform.
Cross-project pairs are constructed, but within-project pairs, which projects actually co-install, show the same drop, as the use of \skillA falls by \dUseAWithin percentage points (\S\ref{sec:rq1}).

\section{Ethical Considerations}
\label{sec:ethics}

We analyze public repositories, and downloaded skills are stored as data and executed only inside isolated containers without credentials (\S\ref{sec:design-runs}).
Before sharing material with human raters, we scanned it for credentials and personal information.
Apart from the two human raters (authors), the study involves no human participants.

\section{Related Work}
\label{sec:related}

\paragraph{Studies of agent skills and skill benchmarks}
Empirical studies of skills have so far examined skills one at a time.
\citet{liu2026skillswild} analyze 31,132 marketplace skills for security vulnerabilities, and \citet{zhang2026reusable} study the defects that keep skills from being reusable in 138,133 \skillmd files.
Closer to our setting, several studies show that more skills can mean worse results.
Agents often fail to load the needed skills once they must search for them~\cite{liu2026skillsrealistic}, and most of the pass-rate drop in larger libraries comes from wrong or missing skill selection~\cite{song2026skillshadowing}.
All of these benchmark studies score a run by whether its task passes or by retrieval accuracy.
A few recent benchmarks check the requirements written in a skill, such as its procedure, its forbidden operations, or its logical relations~\cite{han2026skilluse,shaposhnikov2026framework,chen2026slbench}, but none installs two skills that do the same job, so none can see one skill take over from the other.
Closest to our question, SkillCoach argues that verifier success can mask runs that pick overlapping distractor skills~\cite{zhu2026skillcoach}, \citet{saha2026skillmd} find that agents prefer an adversarially described copy of a skill, and \citet{wang2026safeskillscollide} examine the security risk of co-installed skills.
We instead start from skills installed in public projects, pair each with a real, benign skill that does the same job, and measure what the conflict does to the core functions of the installed skill and whether the user is told.

\paragraph{Tool and skill selection}
MetaTool evaluates whether models decide to use tools and pick the right one, including among similar choices~\cite{huang2024metatool}.
BiasBusters builds groups of functionally equivalent tools and finds that models fixate on one provider or favor tools that appear earlier~\cite{blankenstein2026biasbusters}, and \citet{faghih2025toolpreferences} show that edited descriptions can make a tool receive more than ten times as much usage.
These studies report which tool is selected, whereas a skill injects instructions that change how the whole task is done, so selecting the similar skill can change the result.

\paragraph{Conflicts among co-installed software components}
Software engineering has long studied components that work alone but fail together.
Feature-interaction research examines features whose combination behaves unexpectedly~\cite{calder2003featureinteraction,apel2013fosd}, and studies of package distributions analyze conflicts between packages~\cite{artho2012packageconflicts,vouillon2013coinstall} and between dependency versions~\cite{wang2020watchman}.
Decca grades dependency conflicts by whether the loaded version covers the features that a project uses~\cite{wang2018decca}, much as exclusive core functions do for skills.
Skills add a new kind of interaction, because the model decides which component runs based on natural-language descriptions.

\section{Conclusion}
\label{sec:conclusion}

A similar skill takes one in five runs away from the installed skill and lowers its fidelity, above all on its exclusive core functions, while task completion stays flat.
Where a skill is installed decides which one runs, the final reply rarely says so, and the conflict is decided at the first read of a skill, where a pre-tool hook keeps the core functions of the installed skill.
Because other agents that read the Agent Skills format, such as Codex, Gemini CLI, and GitHub Copilot, also choose skills from names and descriptions, skill evaluations should measure what a skill uniquely contributes, and platforms should guard the first read and show which skill ran.

\section*{Data Availability}
We provide a replication package at \href{https://github.com/ltroin/conflict}{https://github.com/ltroin/conflict}.
It contains the data rules and prompts, the scripts that rebuild the installation lists from repository snapshots, the identifiers of every skill we analyze, the cases of the \nPairs experimental pairs with their core functions and checks, the annotations, the experiment harness, all \nRunsAll run logs, and the judging and analysis scripts.
Skill contents are released where their licenses permit, and otherwise as identifiers with a script that re-fetches them.

\bibliographystyle{ACM-Reference-Format}
\bibliography{references,references-extra}

\end{document}